\documentclass[reprint,aps]{revtex4-1}
\usepackage{graphicx}% Include figure files
\usepackage{dcolumn}% Align table columns on decimal point
\usepackage{bm}% bold math
\usepackage{hyperref}% add hypertext capabilities
\usepackage{subfigure}
\usepackage{amsmath}
\usepackage{url}
\hypersetup{colorlinks=true,linkcolor=black,filecolor=magenta,urlcolor=black,citecolor=blue,}

\begin{document}

\title{Topology Controlled Potts Coarsening}

\author{J. Denholm}\email{j.denholm@strath.ac.uk}

\affiliation{SUPA and Department of Physics, University of Strathclyde, Glasgow G4 0NG, Scotland, United Kingdom}

\author{S. Redner}\email{redner@santafe.edu}
\affiliation{Santa Fe Institute, 1399 Hyde Park Rd, Santa Fe, New Mexico 87501, USA}

\begin{abstract}
  We uncover unusual topological features in the long-time relaxation of the
  $q$-state kinetic Potts ferromagnet on the triangular lattice that is
  instantaneously quenched to zero temperature from a zero-magnetization
  initial state.  For $q=3$, the final state is either: the ground state
  (frequency $\approx 0.75$), a frozen three-hexagon state (frequency
  $\approx 0.16$), a two-stripe state (frequency $\approx 0.09$), or a
  three-stripe state (frequency $<2\times 10^{-4}$).  Other final state
  topologies, such as states with more than 3 hexagons, occur with
  probability $10^{-5}$ or smaller, for $q=3$.  The relaxation to the frozen
  three-hexagon state is governed by a time that scales as $L^2\ln L$.  We
  provide a heuristic argument for this anomalous scaling and present
  additional new features of Potts coarsening on the triangular lattice for
  $q=3$ and for $q>3$.
\end{abstract}

\maketitle

\section{Introduction}

When a ferromagnet with multiple degenerate ground states is quenched from
above to below its critical point, a coarsening domain mosaic emerges in
which distinct phases compete to prevail in the ordering
dynamics~\cite{gunton1982phase,bray2002theory}.  In contradiction to
continuum theories of coarsening, which predict that the ground state is
ultimately reached, the long-time states that persist in discrete spin
systems can be surprisingly rich when the quench is to zero temperature,
$T\!=\!0$.  Such states are actually metastable but infinitely long lived
when this quench to $T\!=\!0$ is instantaneous.  These persistent states may
be static and geometrically simple, such as stripe states in the kinetic
Ising ferromagnet in spatial dimension
$d=2$~\cite{spirin2001fate,spirin2001freezing}.  An unexpected and
simplifying feature of these stripe configurations is that their occurrence
probabilities can be computed exactly in terms of the spanning probabilities
of continuum
percolation~\cite{Barros_2009,Olejarz_2012,Blanchard_2013,Cugliandolo_2016,Blanchard_2017_a,Yu_2017}.

In contrast, for the $d=3$ kinetic Ising ferromagnet, these persistent states
are often topologically complex and
non-stationary~\cite{Olejarz_a_2011,Olejarz_b_2011}.  An even more striking
feature of the $d=3$ Ising ferromagnet is that the probability to reach the
ground state rapidly decreases with $L$ and realizations that do reach the
ground states play an insignificant role for large $L$.  Almost always, the
final state consists of two, and only two, clusters---one spin up and one
spin down.  These two clusters are intertwined so that each cluster typically
has a high genus.  On the surfaces of these clusters, there are a small but
finite fraction of ``blinker'' spins---spins in which three neighbors are in
the spin-up state and three neighbors are in the spin-down state.
Consequently, blinkers can freely flip between the spin-up and spin-down
states with no energy cost.

\begin{figure*}[ht]
  \includegraphics[width=0.9\textwidth]{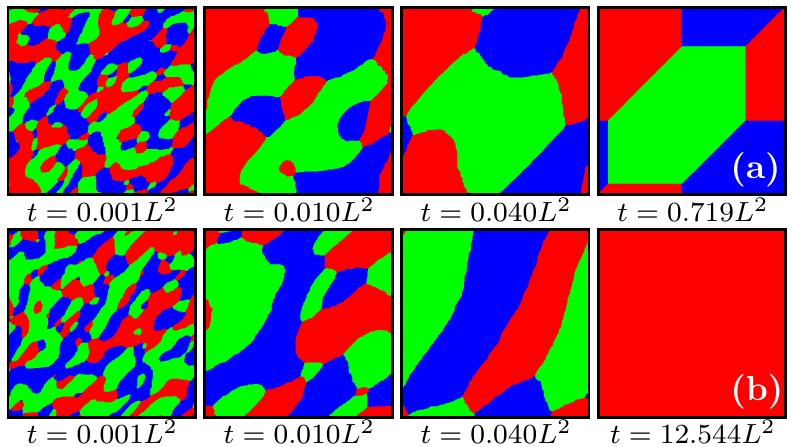}
  \caption{Realizations of zero-temperature coarsening in the $3$-state Potts
    ferromagnet on a periodically bounded triangular lattice of linear
    dimension $L=384$ that freeze into (a) a static three-hexagon state and
    (b) the ground state after evolving through a long-lived off-axis
    three-hexagon configuration.}
  \label{fig:hex}
\end{figure*}

The domain geometry that arises when the kinetic $q$-state Potts ferromagnet
is instantaneously quenched to zero temperature is richer
still~\cite{Safran_1982, Safran_1983, Sahni_1983_a, Sahni_1983_b, Grest_1988,
  Ferrero_2007, Oliveira_2009,LACS10,LAC12}.  The Potts system has been
extensively investigated because of its applications to diverse coarsening
phenomena, such as soap froths~\cite{Glazier_1990, Thomas_2006, Weaire_2009},
magnetic domains~\cite{Srolovitz_1984, Fradkov_1994, Raabe_2000,
  Zollner_2011,Babcock_1990, Jagla_2004}, cellular tissue and other natural
tilings~\cite{Mombach_1990, Mombach_1993, Hocevar2010}.  For $T=0$ quenches,
the ground state is rarely reached for large $q$~\cite{Oliveira_2004_a,
  Oliveira_2004_b} and ``blinker'' (freely flippable) spins arise on the
square lattice~\cite{Olejarz_a_2013}.  A domain mosaic on the square lattice
may also get stuck in a nearly static, geometrically complex state for times
that are much larger than the coarsening time scale.  This metastability is
eventually, but not always, disrupted by a macroscopic avalanche in which
either a lower-energy geometrically complex state or the ground state is
reached~\cite{Olejarz_a_2013}.

In this work, we investigate intriguing and apparently overlooked features of
the coarsening of the 3-state Potts ferromagnet on the \emph{triangular}
lattice.  Our two main results are: (a) When quenched to $T=0$, roughly 75\%
of all trajectories end in the ground state, 16\% in an unexpected frozen
three-hexagon state (Fig.~\ref{fig:hex}(a)), 9\% in a two-stripe state, and a
tiny fraction in a three-stripe state; the probability to reach more complex
geometries, such as frozen states with more than three hexagons, is of the
order of $10^{-5}$.  (b) The approach to the final states is governed by
three distinct time scales: (i) the conventional coarsening time $L^2$, with
$L$ the linear dimension of the system, (ii) a time that appears to grow as
$L^2\ln L$, which governs the approach to frozen three-hexagon states, and
(iii) a time that grows roughly as $L^{3.5}$, which governs the relaxation of
off-axis three-hexagon states or diagonal stripe states to the ground state.
These results will be presented in the following sections.

\section{Triangular 3-State Potts Ferromagnet}

It is convenient to represent the triangular lattice as a periodically
bounded square array with additional diagonal interactions to the upper-right
and lower-left next-nearest neighbors (on the square lattice).  It is worth
mentioning that this periodic system cannot be wrapped onto a two-dimensional
torus.  The Hamiltonian of the system is defined as
\begin{equation}
\mathcal{H} = -2J\sum_{i, j}\big[\delta(s_{i},s_{j})-1\big]\,,
\end{equation}
where $\delta(a,b)$ is the Kronecker delta function, and the sum runs over
all nearest-neighbor spin pairs $i,j$.  In this representation, each
misaligned spin pair contributes $+2J$ to the energy, while each aligned pair
contributes zero.  We choose the coupling strength $J$ to be equal to 1 by
measuring all energies in units of $J$.

We use the following simple $T=0$ single spin-flip dynamics: flip events that
decrease or conserve the systems energy are accepted with probability
$1$~\cite{Sahni_1983_a,Bortz}, while flip events that increase the energy
have zero probability of occurring.  We use an event-driven algorithm to
implement this dynamics in a rejection-free manner~\cite{Sahni_1983_a,Bortz}.
Spins are categorized into classes $k$, that are labeled by the number $F_k$
of distinct permissible flip events that spins in this class may undergo.
The total weight of each class is $W_k = F_kN_k$, where $N_k$ is the number
of spins in the $k^{\rm th}$ class.  To flip a spin, we select a class with a
probability proportional to its weight $W_k$ and allow a randomly chosen
member spin to flip to any energetically allowed spin state with unit
probability.  The time is then incremented by
$\Delta t = -\ln(r)/\sum_k{F_k}$, where $r$ is a uniform random number on the
interval $(0, 1)$, and the summation is over the total number of permissible
flips in the system at the time of the event.  We then update the lists of
spins in each class.

We simulate systems of linear dimension $L$ between 12 and 384, with $10^5$
realizations for each size.  We choose an initial condition that is either a
random zero-magnetization state or an antiferromagnetic state.  Both give
virtually identical results and we henceforth restrict ourselves to the
antiferromagnetic initial condition for simplicity.  In this case, we only
need to average over trajectories of the spin state of the system, rather
than averaging over many spin-state trajectories and also many initial
conditions.

\section{Time Dependence of the Relaxation}

\begin{figure}[ht]
    \centering
    \includegraphics[width=1\columnwidth]{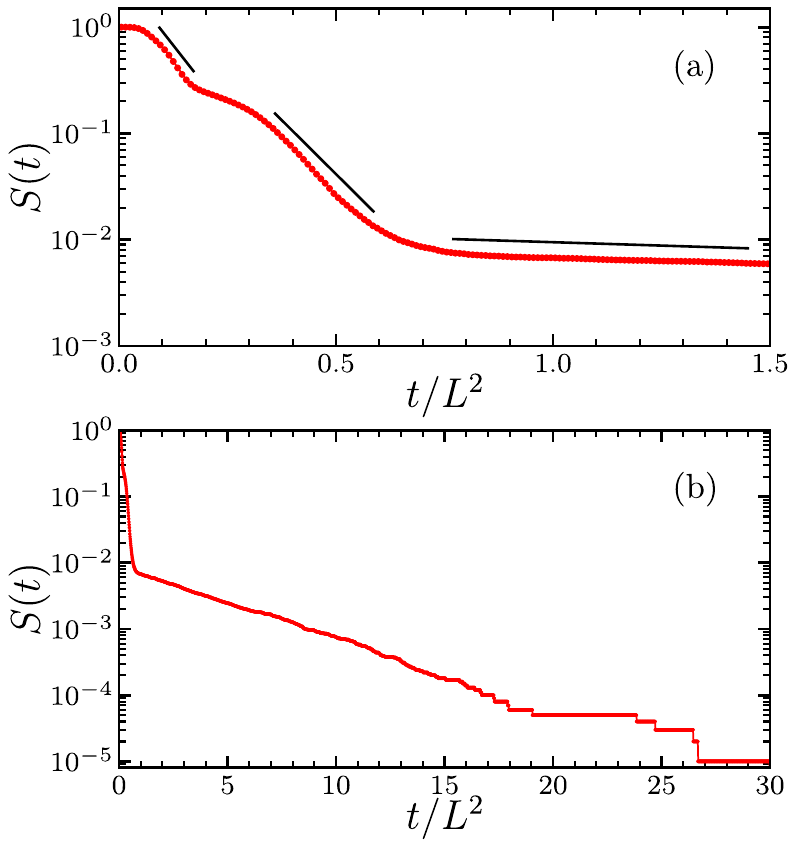}
    \caption{\small Time dependence of the survival probability $S(t)$ for
      (a) $t/L^2\le 1.5$ and (b) $t/L^2\le 30$ for a system of linear
      dimension $L=384$.  In (a), the lines schematically indicate the
      different decay rates associated with coarsening, relaxation to the
      frozen three-hexagon state, and relaxation to the off-axis
      three-hexagon/diagonal stripe states. }
    \label{fig:energy_survival}
\end{figure}

In the conventional picture of phase-ordering kinetics, a finite system of
linear dimension $L$ that is prepared in a random initial state and then
instantaneously quenched to $T=0$ will eventually reach the ground state in a
time that grows with system size as
$L^2$~\cite{gunton1982phase,bray2002theory}.  We therefore expect that the
probability $S(t)$ that the system has not yet reached the ground state at
time $t$, which we define as the ``survival'' probability, will decay
exponentially with time, $S(t)\sim e^{-t/\tau(L)}$, with an associated
relaxation time $\tau(L)$ that grows as $L^2$.  Equivalently, $S(t)$ can be
viewed as probability that flippable spins still exists at time $t$.  Very
different relaxation occurs in the kinetic 3-state Potts ferromagnet on the
triangular lattice.  Here, the time dependence of $S(t)$ appears to be
governed by at least three distinct time scales
(Fig.~\ref{fig:energy_survival}).

For this Potts system, the survival probability decays to zero for all
realizations in a finite time; the longest lifetime in $10^5$ realizations
for $L=384$ is $29.989\,L^2$.  This inertness of all final states also arises
in the kinetic Ising ferromagnet on the square lattice.  A static final state
contrasts with the kinetic Potts ferromagnet on the square lattice, where
blinker spins persist and $S(t)$ never decays to zero~\cite{Olejarz_a_2013}.

\begin{figure}[ht]
    \centering
    \includegraphics[width=1\columnwidth]{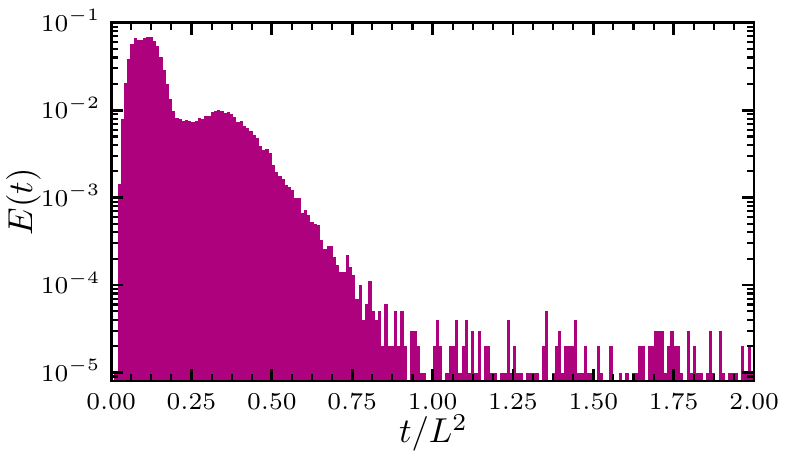}
    \caption{\small Time dependence of the extinction-time distribution
      $E(t)$ for a system of linear dimension $L=384$ with
      $0 \le t / L^{2} \le 1$. }
    \label{fig:extinction-time}
\end{figure}

At short times ($0.05\alt t/L\alt 0.1$), $S(t)$ decays exponentially in time,
with a characteristic decay time that scales as $L^2$, corresponding to
standard coarsening.  This coarsening regime is more readily visible by
studying the probability $E(t)$ that the system goes ``extinct'' at time $t$;
this extinction time corresponds to the time when the last flippable spin
disappears.  This extinction-time distribution is just the negative of the
time derivative of the survival probability.  As shown in
Fig.~\ref{fig:extinction-time}, this distribution has a well-defined
short-time peak whose location increases with $L$ as roughly $L^2$.

At long times, defined by $t/L^2\agt 0.5$, $S(t)$ decays extremely slowly due
to the formation of long-lived diagonal stripe
states~\cite{spirin2001fate,spirin2001freezing} or off-axis three-hexagon
states (one such example is given in the $3^{\rm rd}$ panel of
Fig.~\ref{fig:hex}(b)).  From the asymptote of
Fig.~\ref{fig:energy_survival}(a), we roughly estimate the probability for
the Potts system to fall into either of these states as $5\times 10^{-3}$ for
the largest system that we simulated.  When such states form, a large
fraction of spins on the diagonal interfaces are in zero-energy environments
and thus are freely flippable.  As a result, interfaces that are misaligned
with the lattice axes are able to diffuse.  When two diffusing diagonal
interfaces meet, energy-lowering spin flip events occur in which two disjoint
spin domains merge.  Subsequently, the system quickly falls to the ground
state.

\begin{figure}[ht]
  \centering \includegraphics[width=0.45\textwidth]{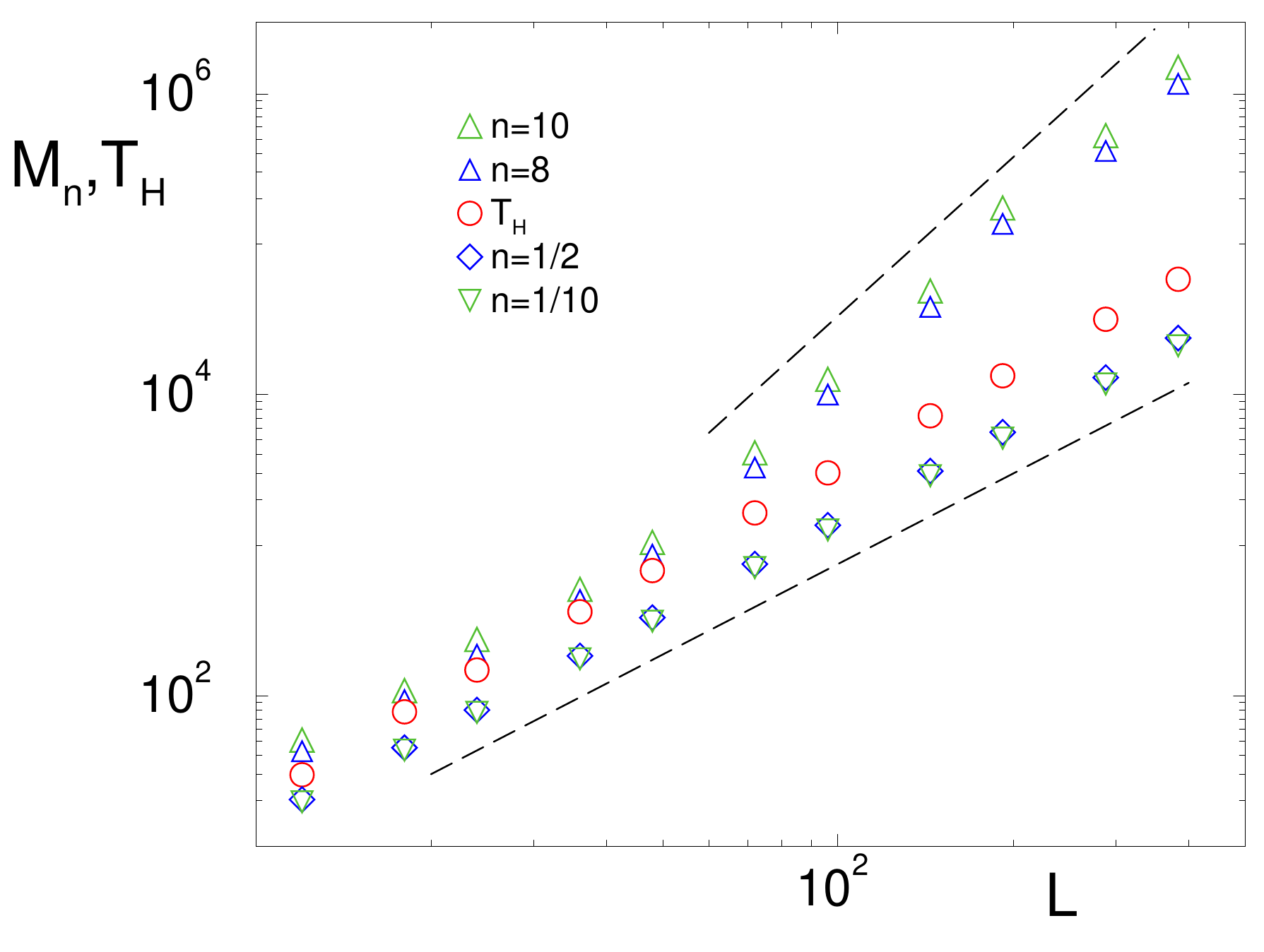}
  \caption{\small The time scales for the kinetic triangular Potts
    ferromagnet: (i) $T_D$, the off-axis three-hexagon/diagonal stripe
    relaxation time, which is obtained from $M_n$ for large $n$ ($n=8$ and
    $n=10$ in the plot).  (ii) $T_H$ ({\color{red}{$\circ$}}), the
    three-hexagon relaxation time.  (iii) The coarsening time, which is
    obtained from $M_n$ for small $n$ ($n=\frac{1}{2}$ and $n=\frac{1}{10}$
    in the plot).  The dashed lines have slopes 3.5 and 2.}
  \label{fig:moments}
\end{figure}

For the analogous diagonal stripe states in the kinetic square-lattice Ising
ferromagnet, we previously argued that this time to reach the ground state
via the diagonal stripe state scales as $L^\mu$ with $\mu=3$ (although
simulation data indicates that this exponent is closer to
3.5)~\cite{spirin2001fate,spirin2001freezing}.  For the Potts ferromagnet, we
find that this corresponding relaxation time, $T_D$, defined as the time for
a system, which enters an off-axis three-hexagon state or a diagonal stripe
state, to eventually reach the ground state, again scales as $L^\mu$, with
$\mu\approx 3.5$ (Fig.~\ref{fig:moments}).

To identify these different time scales, it is helpful to define the reduced
moments of the extinction time distribution
$M_n\equiv \langle t^n\rangle^{1/n}$, with the moment itself defined as
\begin{align}
  \langle t^n\rangle \equiv \int_0^\infty dt\, t^n E(t)\,.
\end{align}
By construction, $M_n$ has the units of time for any $n$ and each $M_n$
defines a characteristic time scale of the coarsening process.  For large
$n$, $M_n$ is dominated by the slowest events in $E(t)$ and we identify these
with the off-axis three-hexagon/diagonal stripe state relaxation time $T_D$.
These high-order moments grow as $L^\mu$, with $\mu= 3.49$ for $M_8$ and
$\mu=3.50$ for $M_{10}$.  Conversely, for small $n$, $M_n$ is dominated by
the fastest events in $E(t)$, which we identify with the usual coarsening
time scale.  These low-order moments grow as $L^\nu$ with $\nu= 2.03$ for
$M_{1/10}$ and $\nu=2.06$ for $M_{1/2}$.

The most interesting dynamics occurs within an intermediate time range
defined by $0.2\alt t/L^2\alt 0.5$.  Here, $S(t)$ decays with time somewhat
more slowly than in the coarsening regime; we argue that this slower time
dependence is a manifestation of the spin system reaching a frozen
three-hexagon state.  We quantify this relaxation by measuring the average
time $T_H$ for the system to reach this three-hexagon state.  As a function
of $L$, a naive power-law fit suggests that $T_H\sim L^\eta$, with
$\eta\approx 2.18$.  However, there is a consistent, but small, downward
curvature in the data of $T_H$ versus $L$ on a double logarithmic scale
(which becomes visible by magnifying Fig.~\ref{fig:moments} and/or viewing
the data for $T_H$ edge on), and a power-law fit is clearly inappropriate.

To help determine the asymptotic behavior of $T_H$, we examine the local
slopes in the plot of $T_H$ versus $L$ that are based on six successive data
points of the eleven data points in all (i.e., between points 1--6, points
2--7, $\dots$, points 6--11).  These local slopes systematically decrease as
the upper limit increases and linearly extrapolate to a value of
approximately 2.1.  This systematic dependence, as well as an exponent close
to an integer value suggests the possibility that $T_H$ might be better
accounted for by the form $T_H\sim L^2\ln L$.  Indeed, a power-law fit of
$T_H/\ln L$ versus $L$ gives a much better fit to the data data, albeit with
an exponent value of 1.93.  However, the data the local exponent based on
successive 6-point slopes now shows a very small upward curvature which
suggests a larger asymptotic exponent value.  Linear extrapolation of the
local slopes gives an exponent estimate of 1.95.  Based on these numerical
results, we are led to the conclusion that $T_H\sim L^2\ln L$.

\begin{figure}[ht]
  \centering \includegraphics[width=0.45\textwidth]{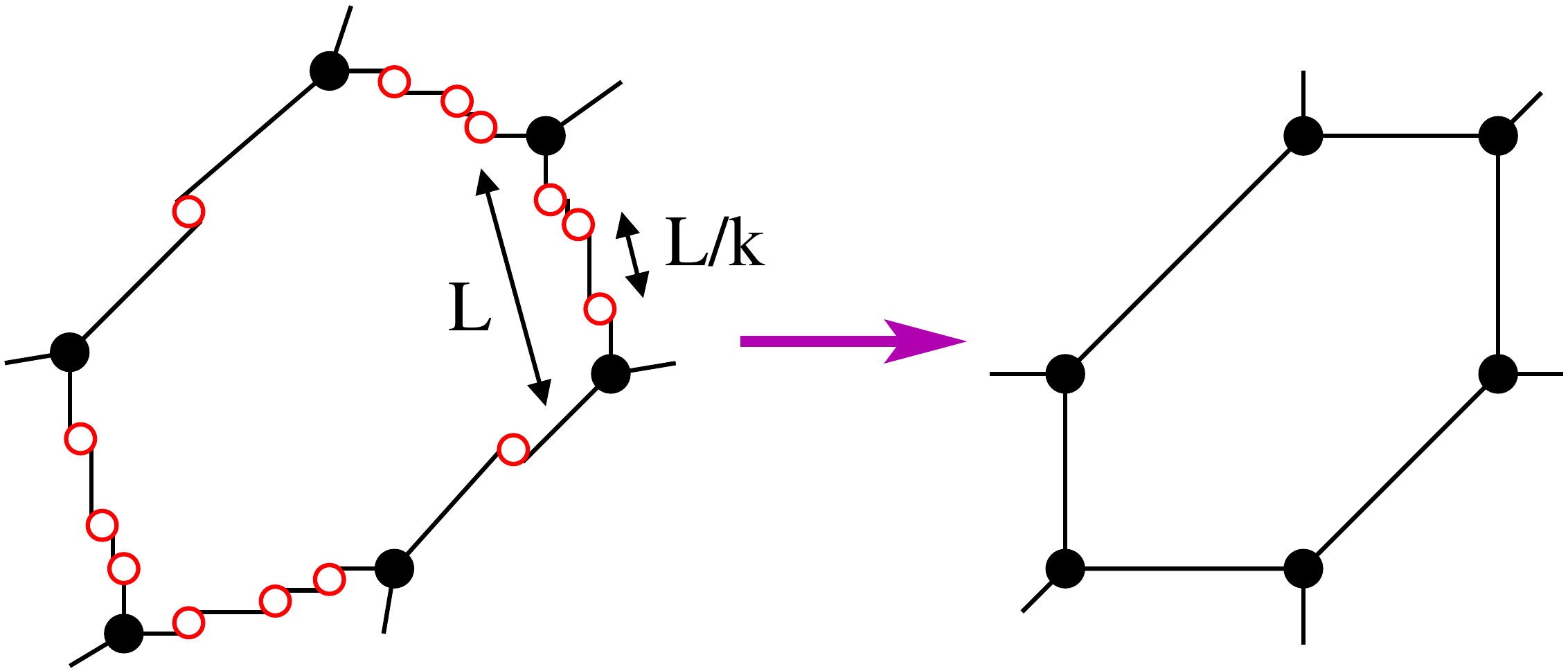}
  \caption{\small Schematic evolution of the evolution of an off-registry
    3-hexagon state to a frozen 3-hexagon state.  The red circles indicate
    freely flippable spins and the heavy dots indicate T-junctions.}
  \label{fig:sketch}
\end{figure}

\begin{figure}[ht]
\centerline{\includegraphics[width=0.25\textwidth]{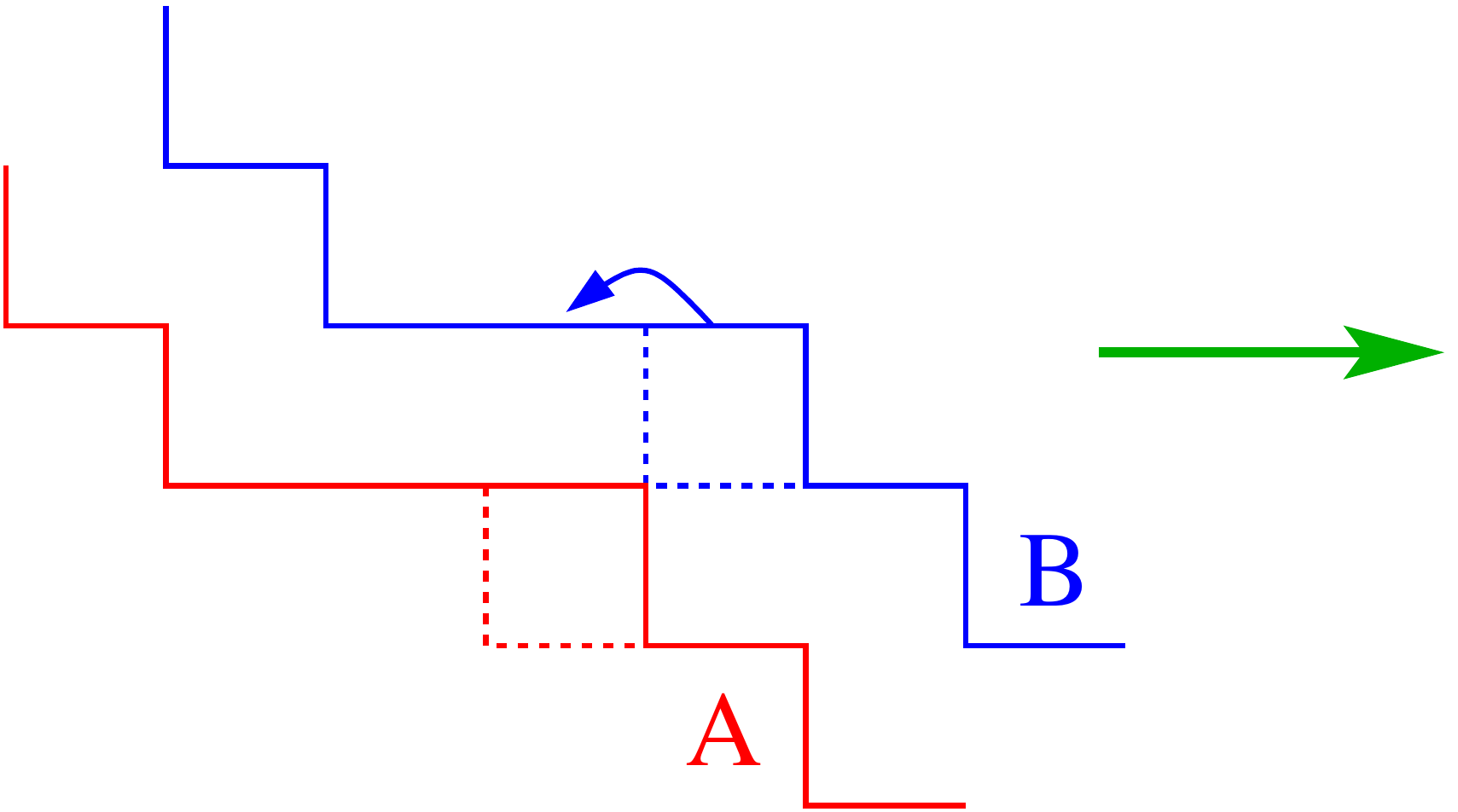}\quad
\includegraphics[width=0.19\textwidth]{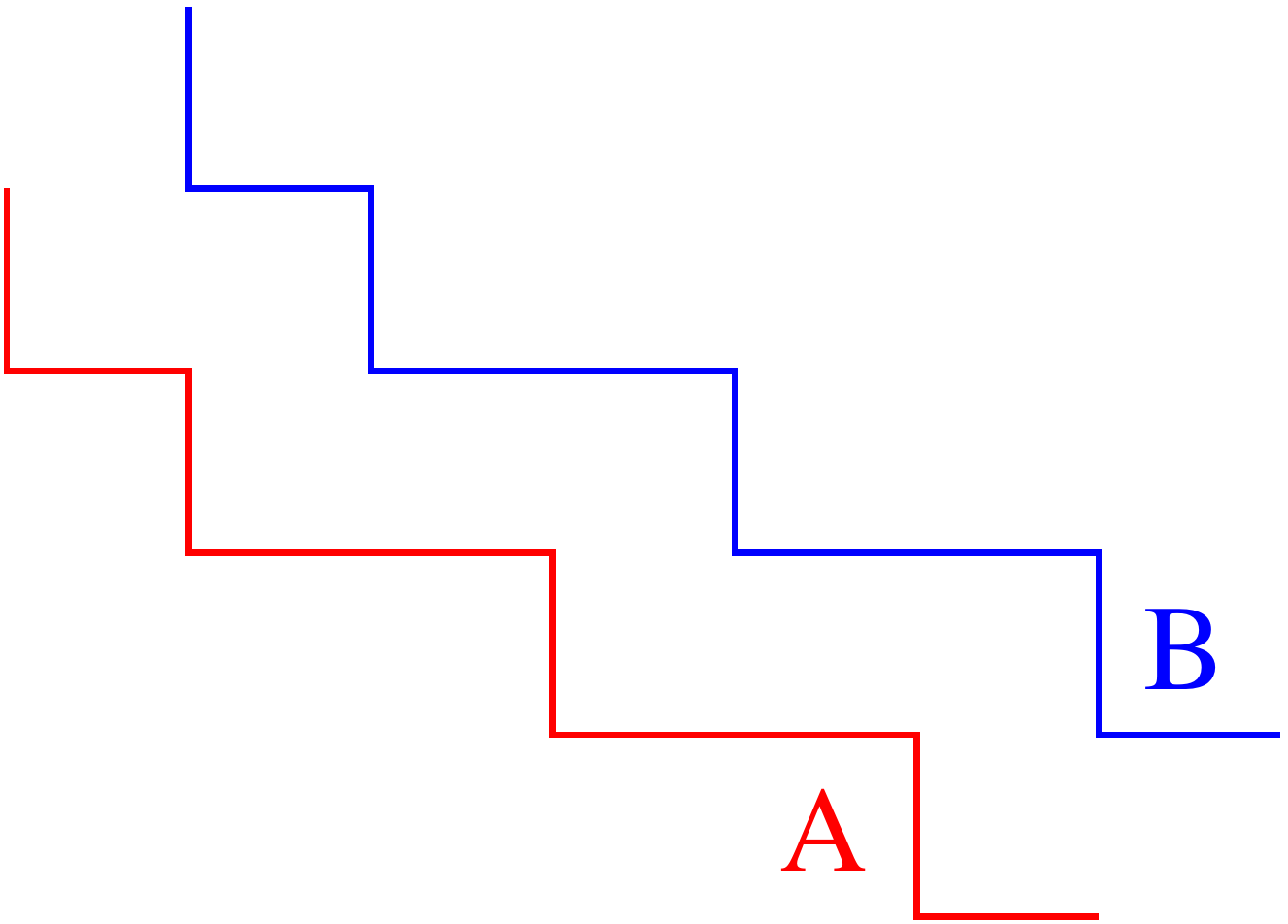}}
\caption{\small (a) A diagonal stripe interface on the triangular lattice.
  The $A$ and $B$ spins on corners can flip with no energy cost.  (b) The
  configuration after a spin flip.}
\label{fig:zigzag}
\end{figure}

This dependence of $T_H$ on $L$ appears to have a simple geometrical origin.
To reach a frozen three-hexagon state, an initial realization first has to
condense to a state that consists of three clusters, none of which span the
system (shown in the $3^{\rm rd}$ panel of Fig.~\ref{fig:hex}(a) and
schematically on the left side of Fig.~\ref{fig:sketch}).  This three-cluster
state contains geometric distortions whereby the six T-junctions---points
where three interfaces meet---are out of registry compared to the aligned
T-junctions in the frozen three-hexagon state ($4^{\rm th}$ panel of
Fig.~\ref{fig:hex}(a)).  Each of the interfaces between pairs of adjacent
T-junctions is thus tilted with respect to a triangular lattice direction.
This means that a substantial fraction of the spins on each such interface
are freely flippable.  As indicated in Fig.~\ref{fig:zigzag}, each freely
flippable spin on an interface is equivalent to an independent random walker
that can hop along the interface~\cite{KRB10}.

The tilted interfaces must gradually straighten for the configuration to
reach the frozen three-hexagon state~\cite{movie}.  This straightening
process occurs by the motion of the equivalent random walkers.  When a random
walker reaches a T-junction, the position of the latter moves by one lattice
spacing.  This displacement corresponds to the random walker being absorbed
at the T-junction.  Thus we can view the process of interface straightening
as equivalent to the successive absorption of the order of $L$ independent
random walkers on a finite interval whose length is also of the order of $L$.

When there are $k$ walkers in an interval, their typical separation is $L/k$;
this is also the distance between the end of the interval and the closest
walker to the interval end.  The first-passage time until this closest walker
reaches the end of the interval and is absorbed there is given by
$t_k=\frac{L}{k}\big(L-\frac{L}{k}\big)$~\cite{redner2001guide}.  When all
the walkers along the interfaces have been absorbed, the final, frozen
three-hexagon state has been reached.  By adding these individual absorption
times until all walkers have been absorbed, the time to reach the frozen
three-hexagon state is (ignoring constants of order 1),
\begin{align}
  \label{t-hex}
  \tau = t_L+t_{L-1}+\ldots + t_1%\nonumber\\
      &= \sum_{k=1}^L\frac{L}{k}\Big(L-\frac{L}{k}\Big)\nonumber\\[1mm]
       &  \simeq L^2\ln L\,.
\end{align}
While our argument is crude, it appears to capture the mechanism that
underlies the approach to the frozen three-hexagon state.  Our prediction is
consistent with the simulation results shown in Fig.~\ref{fig:moments}.

\section{Final States}

\begin{figure}[ht]
    \centering
    \includegraphics[width=0.9\columnwidth]{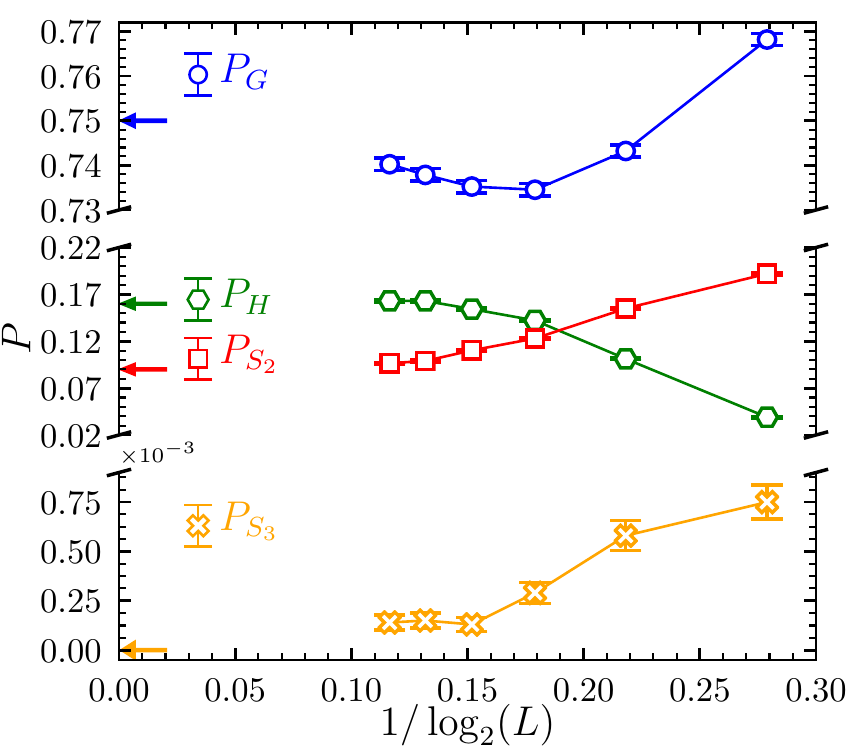}
    \caption{\small Probabilities of freezing into the ground state, $P_G$, a
      two-stripe state, $P_{S_{2}}$, a frozen three-hexagon state, $P_H$, and
      a three--stripe state $P_{S_{3}}$ as a function of $1/\log_2 L$.  Our
      $L\to\infty$ estimates of these probabilities are 0.75, 0.09, 0.16, and
      0 respectively (arrows).}
    \label{fig:estimated_probabilities}
\end{figure}

A striking aspect of the coarsening of the 3-state triangular Potts
ferromagnet is that a new type of final state---a configuration that consists
of three hexagons---is reached with a non-zero probability for $L\to\infty$.
Figure~\ref{fig:estimated_probabilities} shows the $L$ dependence of the
probabilities for the system to eventually reach: the ground state
(probability close to 0.75), a frozen three-hexagon state (probability close
to 0.16), a two-stripe state (probability close to 0.09), and three-stripe
state (with probability of the order of $10^{-4}$) for the largest system
simulated.  The three-stripe state plays a negligible role in the coarsening
dynamics.  Because of the non-monotonic and/or slow $L$ dependences of the
final-state probabilities, our estimates for their $L\to\infty$ values are
necessarily crude.  Similar extrapolation issues were encountered in the
kinetic Ising ferromagnet and the square-lattice Potts
ferromagnet~\cite{spirin2001fate,spirin2001freezing,Olejarz_a_2013}.

\begin{figure}[ht]
    \centering
    \includegraphics[width=0.9\columnwidth]{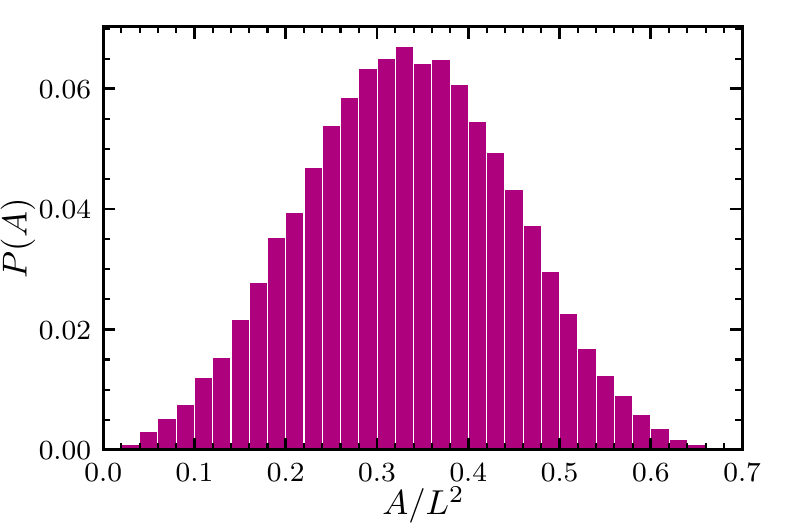}
    \caption{\small Distribution of the domains areas in the three-hexagon
      final states for a system of linear dimension $L=384$.}
    \label{fig:area}
\end{figure}

Intriguingly, the energy of any three-hexagon state (such as the example
shown in Fig.~\ref{fig:hex}(a)) equals $24L$, independent of the individual
hexagon sizes.  By examining the $4^{\rm th}$ panel of Fig.~\ref{fig:hex}(a),
the total length of each of the vertical, horizontal, and tilted interfaces
in this state must equal $L$.  Since there are two spins in different states
on either side of the interface, there are $6L$ interfacial spins in total.
Because an interfacial spin has four neighbors in the same state and two
neighbors in a different state, each such spin contributes $+4J$ to the total
energy.  Consequently, the final energy of any frozen three-hexagon state is
$24L$.  Although the total perimeter of the three-hexagon state is fixed, the
area of each hexagon is a random quantity whose distribution has a
well-defined peak near $\frac{1}{3}$ (Fig.~\ref{fig:area}).  This behavior
visually mirrors what was found previously in the kinetic Ising ferromagnet.
Here, roughly 1/3 of all realizations condensed into a stripe state, in which
the width distribution of the stripes was reasonably fit by a Gaussian
distribution~~\cite{spirin2001fate,spirin2001freezing}.

\begin{figure}[ht]
\centerline{\includegraphics[width=0.3\textwidth]{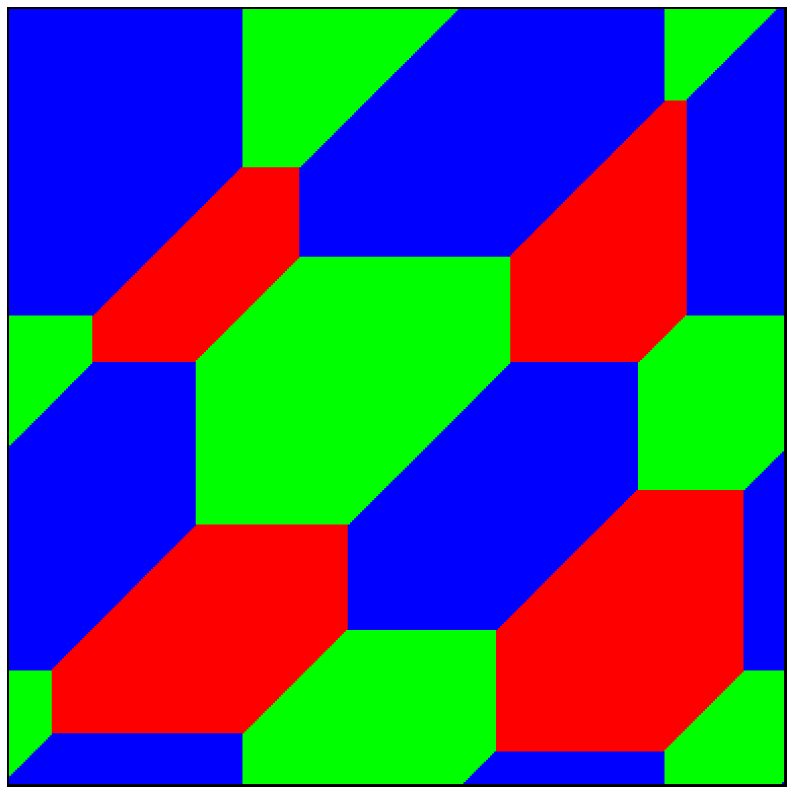}}
\caption{\small A twelve-hexagon final state in the $q=3$-state Potts
  ferromagnet.  This configuration was found to occur once in $10^5$
  realizations for a system of linear dimension $L = 384$.}
  \label{fig:q3-hex12}
\end{figure}

Finally, for the 3-state Potts ferromagnet, we also observe static final
states that contain more than three hexagons with a vanishingly small
probability.  Shown in Fig.~\ref{fig:q3-hex12} is an example of a
twelve-hexagon state that was observed once in an ensemble of $10^5$
realizations for a system of linear dimension $L=384$.  Using the same
reasoning as that given for the three-hexagon state, it is straightforward to
infer that the energy of this twelve-hexagon final state is $48L$.
Intriguingly, we did not see, static states that consist of six hexagons in
$10^5$ realizations.  In hindsight, six-hexagon states should not appear
because such states cannot be symmetrically situated within a finite-size
square domain.

\section{Potts Ferromagnet with  $q>3$ States }

\begin{figure}[ht]
\includegraphics[width=0.5\textwidth]{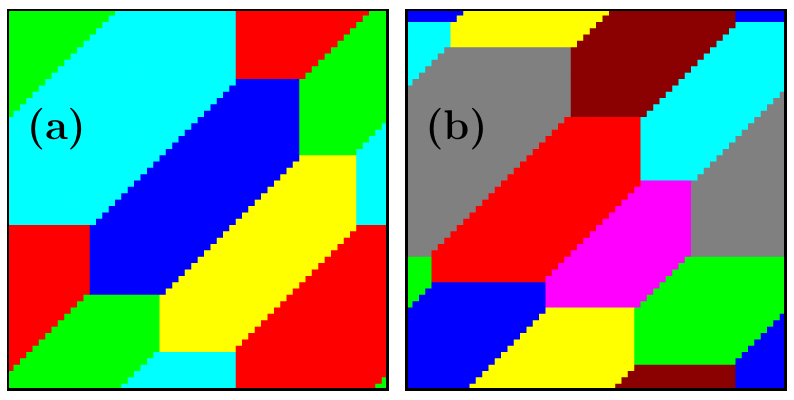}
\caption{\small (a) A five-hexagon final state of the $q=6$-state Potts
  system, and (b) an eight-hexagon state in the $q=20$ Potts system.}
  \label{fig:q5-6}
\end{figure}

Given the rich dynamical behavior of the 3-state Potts ferromagnet, it is
natural to investigate this same model with more than three spin states.  The
dynamics and long-time states of the $q>3$ system shares many features with
the 3-state Potts ferromagnet, but additional unusual feature arise.  As the
number of spin states is increased, the coarsening mosaic becomes visually
more picturesque and the possible final states are correspondingly more
complex~\cite{Safran_1982, Safran_1983, Sahni_1983_a, Sahni_1983_b,
  Grest_1988, Ferrero_2007, Oliveira_2009}.  Final states that contain more
than three hexagons now arise with non-negligible probabilities.  To give
some examples, for $q=6$ and $L=60$ and 120, we observed five-hexagon states
99 and 106 times, respectively, out of $10^{5}$ realizations
(Fig.~\ref{fig:q5-6}(a)).  For $q=20$ and $L=60$ and 120, five-hexagon states
were observed 215 and 165 times, respectively, out of $10^5$ realizations.
We also observed 7 eight-hexagon states out of $10^{5}$ realizations for
$q=20$ and $L=60$, but did not observe any such states for L=120
(Fig.~\ref{fig:q5-6}(b)).

Final states that contain blinker spins also exist, but these are extremely
rare.  We observed blinker spins for $q=5$ and $q=6$ states with a
probability of the order of $10^{-4}$, but only for small system sizes.  We
did not observe blinker spins in any triangular Potts ferromagnet with
$L>40$.  Both of these exotic long-time states---multi-hexagon states and
blinker spins---occur sufficiently rarely that they play a negligible role in
characterizing the coarsening dynamics.

\begin{figure}[ht]
  \centering \includegraphics[width=0.5\textwidth]{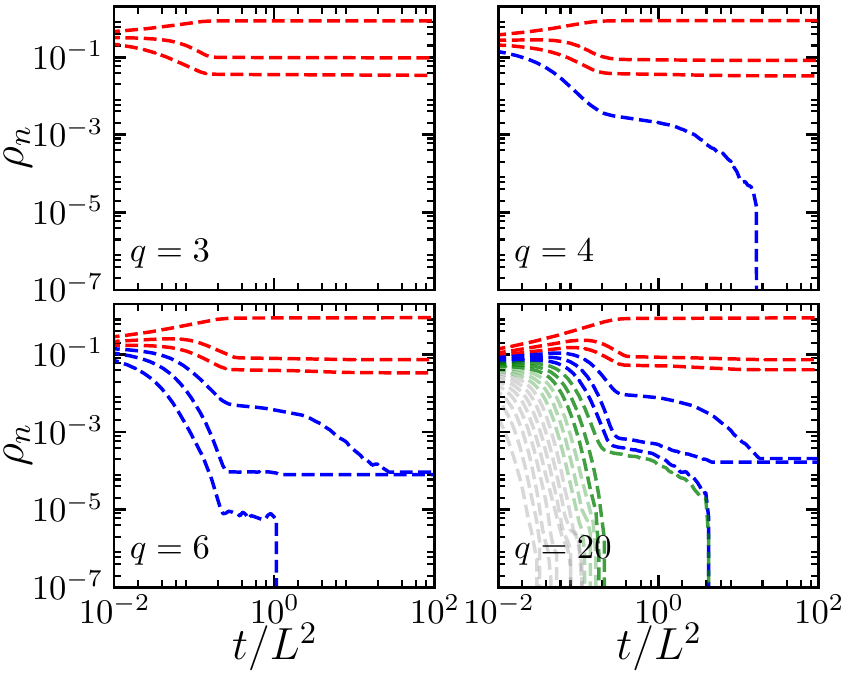}
  \caption{\small Time evolution of the densities of each spin type,
    $\rho_n$, sorted by abundance order.  The data is based on $10^{4}$
    realizations on systems of linear dimension $L=240$.}
  \label{fig:q_populations}
\end{figure}

Another intriguing aspect of the large-$q$ Potts ferromagnet is the near
universality of the long-time densities of the most-common spin type, the
second most-common type, etc., (Fig.~\ref{fig:q_populations}).  Let us denote
by $\rho_1$, the fraction of the most-common spin type in the final state,
$\rho_2$, the second most-common spin fraction, etc.  Starting with the
antiferromagnetic state, with equal numbers of each spin type, the final
fractions of the three most abundant spin types are
($\rho_1,\rho_2,\rho_3)\approx (0.870,0.096,0.034)$ for $q=3$ and
$(0.893,0.073,0.036)$ for $q=6$ (red in Fig.~\ref{fig:q_populations}).  For
$q$ between 3 and 6, the fraction of spins types outside the top three
abundances is less than $2\times 10^{-4}$.  For $q>6$, the final fractions
$\rho_n$ for the five most abundant spin types are nearly universal, while
the final fractions $\rho_n$ for $n>5$ are negligibly small.  Thus
simulations of Potts ferromagnets with $q>6$ will not reveal new long-time
physical features compared to Potts ferromagnets with $q\leq 6$.  It is
possible there could be final states that contain richer arrangements of
hexagons, but these states would play a negligible role in understanding the
overall coarsening process.

\section{Concluding Remarks}

To summarize, the kinetic $q$-state Potts ferromagnet on the triangular
lattice exhibits a variety of intriguing topological features.  For $q=3$,
the final configurations are all static and either: the ground state, frozen
three-hexagon states, two-stripe states, or three-stripe states, with
respective frequencies of 75\%, 9\%, 16\%, and $<$ 0.02\%.  Frozen final
states that contain more than three hexagons occur with a probability that is
less than $10^{-5}$.  The dynamics is governed by three distinct time scales:
a coarsening time that grows as $L^2$, a hexagonal state condensation time
$T_H$, and an off-axis hexagon/stripe condensation time that grows roughly as
$L^{3.5}$.  We argued, based on mapping freely flippable spins on hexagonal
domain interfaces to a set of independent absorbing random walkers in a
finite interval, that $T_H\sim L^2\ln L$, a prediction that is consistent
with simulation data.

We also found that the dynamical behavior of $q>3$-state Potts ferromagnet on
the triangular lattice is not materially different than that of the 3-state
Potts ferromagnet.  For $q>3$ and when the initial state is
antiferromagnetic, only the three most abundant spin types are present in
measurable amounts at long times.  A new feature of the final states for
$q>3$ is that frozen configurations that contain more than three hexagons
arise.  The occurrence probability for these exotic configurations is much
larger than in the $q=3$ Potts system, but still only of the order of
$10^{-3}$.

There are a variety of open questions raised by this work.  First, is it
possible to compute the probability to reach the frozen three-hexagon state?
The percolation mapping proved decisive to understand the occurrence of
various stripe topologies in the kinetic Ising
ferromagnet~\cite{Barros_2009,Olejarz_2012,Blanchard_2013,Cugliandolo_2016,
  Blanchard_2017_a,Yu_2017}.  Perhaps there is a mapping between final states
of the 3-state Potts ferromagnet and the 3-color percolation model, which has
only begun to be investigated~\cite{Z77,SY14}.  A second open question is to
understand the area or the perimeter distribution of the three-hexagon state.

The fact that the final states are simply categorized on the triangular
lattice also raises the question of whether there are simple final states for
the 3-state Potts ferromagnet on other 6-coordinated lattices, such as the
simple cubic lattice.  Perhaps there is underlying simplicity when the
lattice coordination number is an integer multiple of the number of Potts
states.  Another unresolved question is the characterization of the final
states of the 3-state Potts ferromagnet on the square lattice.  While these
final states are visually rich and many are apparently
non-static~\cite{Olejarz_a_2013}, they have yet to be quantitatively
characterized.

We thank Ben Hourahine and Cris Moore for helpful discussions.  JD thanks
EPSRC DTA5 grant EP/N509760/1, the Mac Robertson Trust, and the Santa Fe
Institute.  SR thanks support from NSF grant DMR-1608211.  We acknowledge
ARCHIE-WeSt High Performance Computer based at the University of Strathclyde
as well as grant EP/P015719/1 for computer resources.

%\bibliography{references.bib}
%merlin.mbs apsrev4-1.bst 2010-07-25 4.21a (PWD, AO, DPC) hacked
%Control: key (0)
%Control: author (8) initials jnrlst
%Control: editor formatted (1) identically to author
%Control: production of article title (-1) disabled
%Control: page (0) single
%Control: year (1) truncated
%Control: production of eprint (0) enabled
%

\end{document}